**Tractor Service Utilisation, Profitability, and Adoption Determinants Among Smallholder Maize Farmers in Ejura-Sekyedumase Municipality, Ghana**

**Innocent Yao Yevu[a,*], Fred Nimoh[a], Edward Arthur[a], Attah-Nyame Essampong[a], Asante Emmanuel Addo[a] and Addai Kevin[a]**

[a]*Department of Agricultural Economics, Agribusiness and Extension, Kwame Nkrumah University of Science and Technology, Kumasi, Ghana*

* *Correspondence: innocentbrianne01@gmail.com*

**Abstract:** Although tractor services are increasingly used in the Ejura-Sekyedumase Municipality, access remains uneven, and some smallholder maize farmers still rely on labour-intensive production methods. This study investigated the profitability effects and adoption determinants of tractor service utilisation in Ejura-Sekyedumase Municipality, Ashanti Region, Ghana. Cross-sectional data were collected from 359 farmers using multi-stage proportionate random sampling. A multivariate probit model identified simultaneous adoption determinants across ploughing and shelling services. Propensity score matching (PSM) estimated a positive profitability effect of GH₵471 per acre after controlling for observable selection bias. Users achieved a net profit margin of 29.87% and a return-to-cost ratio of 42.60%, compared with 25.10% and 33.51% for non-users, respectively. Farming experience and fertiliser intensity positively predicted adoption, while farm size exerted a consistent negative influence; this may suggest supply-side availability constraints, including potential fleet-capacity limitations. The negative FBO result may reflect labour-sharing functions that substitute for paid mechanized services; this mechanism should be tested directly in future work. Kendall's W (0.381) indicated moderate agreement in farmers' rankings, with financial constraints ranked as the leading barrier. The study recommends targeted credit facilitation, possible AMSEC fleet expansion pending supply-side verification, intensified extension services, gender-responsive mechanisation policies, and redesigned FBO programmes.

## 1. Introduction

Agricultural mechanisation is increasingly recognised as a critical driver of productivity, income improvement, and food security among smallholder farmers in sub-Saharan Africa (SSA) (Mdoda et al., 2022; Takeshima et al., 2013; Owusu et al., 2011). Despite its transformative potential, SSA averages fewer than two tractors per 1,000 hectares of arable land, compared with approximately ten tractors per 1,000 hectares in South Asia and Latin America (FAO, 2022), while approximately 70% of SSA farmers cultivate plots smaller than two hectares using hand hoes (Sihlobo et al., 2022), perpetuating a cycle of low productivity, high post-harvest losses, and reduced market competitiveness. Agriculture constitutes a fundamental pillar of Ghana's economy, contributing approximately 20% of GDP and providing livelihoods for approximately 35% of the national workforce (World Bank, 2024). Maize (Zea mays L.) is Ghana's most important cereal crop, accounting for over 50% of total cereal production and cultivated across all agro-ecological zones (Asamoah et al., 2024). Yet production remains predominantly smallholder-based; approximately 80% of total agricultural output is produced by small-scale, family-operated farms employing rudimentary, labour-intensive technologies (Kansanga et al., 2019), perpetuating persistently low yields, income insecurity, and rural poverty.

Government policy in Ghana has progressively advanced mechanisation through the establishment of Agricultural Mechanisation Service Enterprise Centres (AMSECs) and subsidised tractor importation to promote service accessibility (Houssou et al., 2013; Diao et al., 2014; Benin, 2015). However, access to mechanisation services remains constrained by several structural factors. Cossar (2019) found that low utilisation rates render most AMSECs financially unviable even with government subsidies, while Daum & Birner (2017) demonstrated that transaction costs discourage machine operators from serving all categories of farmers, meaning demand does not automatically translate into access. Diao et al. (2014) argued that supply-side gaps now represent the binding constraints in Ghana's mechanisation market, despite growing demand, a finding corroborated by Houssou et al. (2013), who documented inadequate post-harvest mechanisation services as a particular bottleneck. On the demand side, financial constraints are consistently identified as the primary adoption barrier (Tilahun et al., 2024), while insecure land tenure and limited credit jointly suppress uptake in northern Ghana (Lambrecht & Asare, 2016). These interacting supply- and demand-side barriers create a mechanisation paradox in which demand exists, but adoption remains limited.

Within Ejura-Sekyedumase Municipality, one of Ghana's most commercially active maize-producing areas, prior studies have examined mechanisation status (Taiwo & Kumi, 2015), climate variability effects (Cudjoe et al., 2021), maize production failure (Obour et al., 2022), tractor maintenance practices (Aikins et al., 2016), and technical efficiency (Bempomaa, 2014). Nkegbe (2013) found that in northern Ghana, AMSEC tractor access was a significant positive factor influencing the adoption of mechanised services among smallholder farmers. Anang & Asante (2020) also found that farming experience positively predicts access to tractor services. Despite this growing body of evidence, these studies share a common limitation: they focus predominantly on efficiency and input-use outcomes while neglecting the profitability dimension of tractor services. Limited evidence has quantified the adjusted profitability differential between tractor service users and non-users, nor examined adoption determinants simultaneously across multiple service types. This represents a critical knowledge gap, because understanding what drives adoption and whether it translates into measurable profit gains is essential for designing interventions that move beyond productivity metrics toward genuine income improvements for smallholder households. Furthermore, the study by Bempomaa (2014) provides only aggregate constraints rankings without disaggregating by gender or farm size, thereby masking differential barriers that carry distinct policy implications. Moreover, evidence confirms that gender disparities in access to mechanisation are both real and systematic (Dwomoh et al., 2023; Slavchevska et al., 2021).

Propensity score matching (PSM) (Rosenbaum & Rubin, 1983) and the multivariate probit (MVP) framework (Wooldridge, 2016) address these gaps directly. PSM, introduced by Rosenbaum & Rubin and extended in agricultural economics applications by Caliendo & Kopeinig (2008), constructs a comparable non-user group based on observed characteristics and reduces observable selection bias. Recent PSM applications in related mechanisation and agricultural-technology studies, including Mdoda et al. (2022) and Amankwah (2023), show that selection bias can inflate simple mean comparisons. The MVP framework is necessary because ploughing and shelling adoption decisions are non-mutually exclusive, and their error terms may be correlated, conditions that separate binary probit models would misrepresent (Amankwah, 2023; Wooldridge, 2016). Amankwah (2023) demonstrated cross-adoption complementarities in technology adoption, finding that farmers who adopt fertiliser are significantly more likely also to adopt mechanised land preparation, a finding directly relevant to this study's fertiliser-adoption relationship.

The theoretical basis of this study integrates three complementary frameworks. Profit maximization theory (Varian, 2014), rooted in neoclassical microeconomics, predicts adoption when expected net returns, through higher yields, reduced labour costs, and improved input efficiency, exceed the additional cost of mechanised services. Utility maximisation theory (Mas-Colell et al., 1995) complements this by recognising that farm household decisions are shaped by a broader preference function encompassing labour reduction, timeliness, risk management, and social considerations, which helps explain why household size negatively predicts adoption, as larger households substitute family labour for hired machinery. Rogers' (2003) Diffusion of Innovations Theory provides a further lens, linking farming experience and education to perceived compatibility and relative advantage, while financial constraints reflect barriers to trialability. Together, these frameworks ground the variable selection and interpretation of findings in established economic and sociological theory.

For many smallholder farmers, tractor service utilisation is not merely a production decision; it is a livelihood strategy with significant implications for farm yields, input use, and household welfare. Farmers cannot rely solely on manual production methods because labour-intensive practices perpetuate high costs and low returns; the absence of mechanised systems forces smallholders to rely on traditional methods that are less productive and result in lower yields and diminished income (Ansah et al., 2024; Mwangi & Kariuki, 2015); hence, the adoption of mechanised services becomes important to the farmer's profitability and survival. However, it remains unclear whether mechanisation strategies alone can address the profitability challenges of smallholder farmers, or whether this can only be achieved through complementary investments in credit access, extension services, and input supply. Given these uncertainties, this study addressed two concerns.

First, the study analysed the adjusted effect of tractor service adoption on smallholder maize farmers' profitability, using PSM to correct for observable selection bias. This analysis provided empirical evidence on the role of tractor service adoption in improving the profitability and livelihood outcomes of rural smallholder farm households. Second, the joint determinants of tractor service adoption for both ploughing and shelling services were analysed using MVP. Additionally, adoption constraints were further disaggregated by gender and farm size using Kendall's coefficient of concordance, generating distributional insights that aggregate analysis would otherwise mask. The following null hypotheses were formulated to guide the empirical analysis:

1. The adoption of tractor services has no significant effect on the profitability of smallholder maize farmers in Ejura-Sekyedumase Municipality, Ghana.
2. Farmer-level socioeconomic attributes, institutional affiliations, and farm-level characteristics have no significant effect on the adoption of tractor services for ploughing and shelling.

Examining these hypotheses is essential for moving beyond the generalised promotion of mechanisation as a universal solution for improving farm profitability among rural households, without adequately recognising the structural differences that distinguish farmers by gender and farm size. Policies that ignore these heterogeneities risk being poorly targeted and ultimately ineffective for the most constrained farming households. The findings from this study are therefore expected to assist the Ministry of Food and Agriculture (MoFA), development partners, and other stakeholders within Ghana's agricultural sector to better appreciate the diverse mechanisation conditions facing smallholder maize farmers and to craft evidence-based, context-sensitive policies that genuinely improve the profitability and economic well-being of rural farm households.

## 2. Materials and Methods

### 2.1 Study Area

This study was carried out across eight communities within the Ejura-Sekyedumase Municipality, Ashanti Region of Ghana. The municipality is geographically positioned between longitudes 1°5'W and 1°39'W and latitudes 7°9'N and 7°36'N, encompassing a total land area of about 1,782.2 $km^2$. It is situated within the transitional belt separating the forest and savanna agroecological zones of mid-Ghana, a classification that aligns with Zone C of the Ghana Meteorological Agency's (GMet) agroecological zoning framework (Obour et al., 2022). The eight study communities are: Aframso, Dromankuma, Babaso, Teacherkrom, Kasei, Anyinasu, Sekyedumase, and Bayere-Nkwanta (Figure 1). These communities were selected in consultation with the Municipal Department of Agriculture to guarantee comprehensive geographical representation of the municipality's principal maize-producing zones.

The vegetation of the municipality is dominated by semi-deciduous forest cover, gradually giving way to savanna towards the north (Obour et al., 2022). Total annual rainfall across the municipality ranges from 1,200 to 1,500 mm. The first rainy period begins in late March or early April and continues until mid-July. A transient dry spell follows between mid-July and mid-August, after which a shorter second rainy period extends through September and October. From November to March, the municipality enters a prolonged dry phase characterised by the Harmattan, dry and dusty north-easterly trade wind that significantly reduces atmospheric moisture. The bimodal rainfall structure is particularly advantageous for smallholder farmers, as it permits two planting cycles each year, most notably for cereal and legume crops cultivated under rain-fed conditions prevalent across the transitional zone.

With regards to temperature, the municipality records mean monthly values ranging from 21°C to 30°C. The period spanning January through April is typically the warmest, while July and August represent the coolest months of the year. Relative humidity levels are strongly influenced by seasonal rainfall patterns, reaching approximately 90% in June at the height of the wet season and declining to approximately 55% in February during the dry period. These climatic attributes, combined with fertile transitional soils and the municipality's prominence in commercial maize production within the Ashanti Region, collectively justified its selection as the study area.

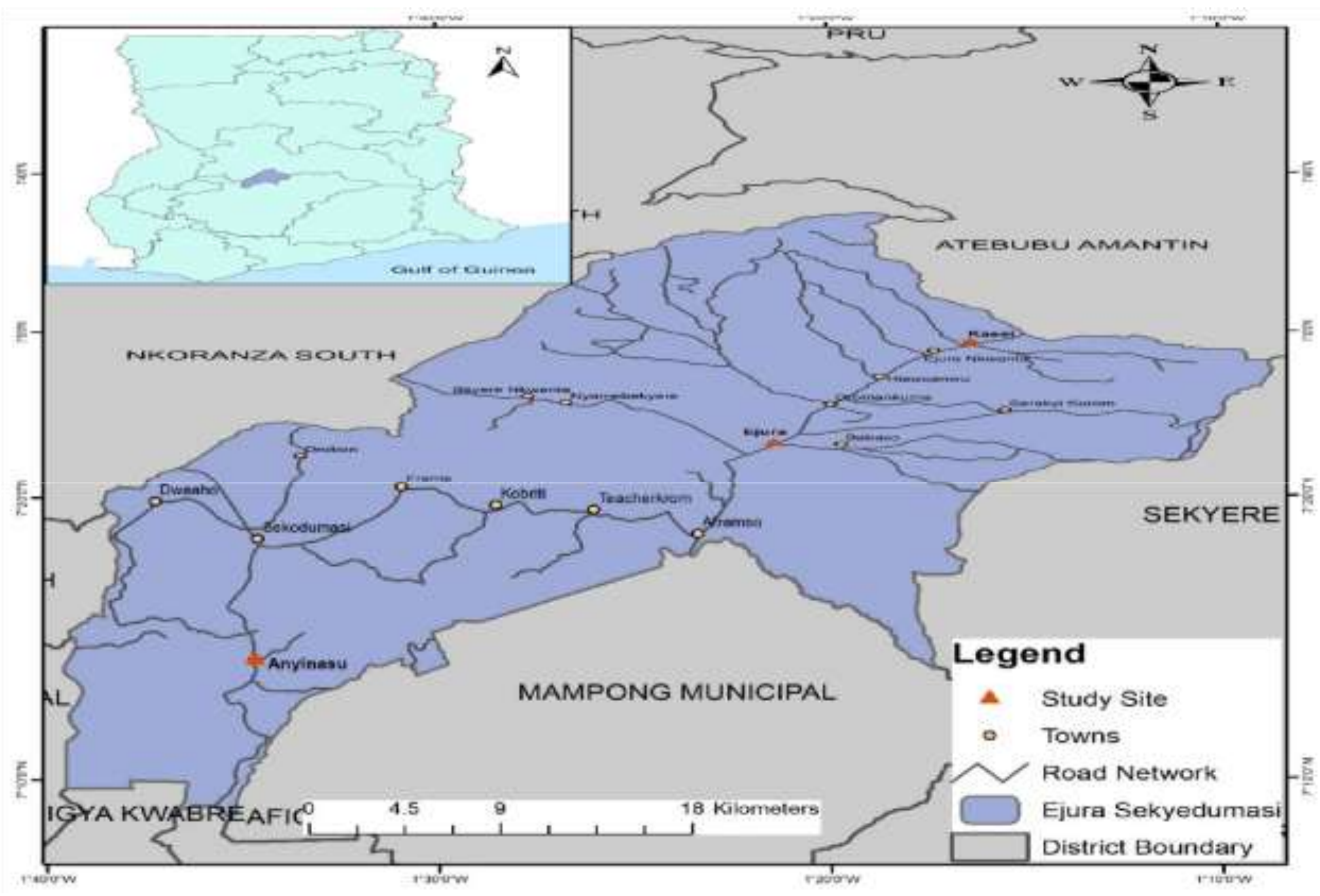

**Figure 1**. Map showing the study communities. Source: Obour et al., 2022.

### 2.2 Research Design, Sampling, and Data Collection

The study employed a positivist approach, using a cross-sectional quantitative research design (Saunders et al., 2019; Creswell & Creswell, 2018). The positivist philosophical tradition holds that social phenomena, including farmers' technology adoption decisions and their economic outcomes, can be measured objectively and subjected to empirical testing. Following the deductive approach, testable hypotheses were derived from established economic theory and subsequently tested against empirical field data. The mono-method quantitative design employed a multi-stage proportionate random sampling to ensure municipal representativeness.

The multistage sampling procedure was used, involving purposive selection of communities followed by proportionate random selection of farmers. In the first stage, eight communities were purposively selected based on the scale of maize production and their recorded use of tractor-hiring services. This selection was intended to capture the geographic diversity and varying levels of tractor service accessibility across the municipality. In the second stage, farmers within each community were selected using proportionate random sampling, whereby the number of respondents allocated to each community was proportional to its estimated farmer population. This approach ensured equitable representation and statistical comparability across the sampled communities. Table 1 presents the distribution of respondents across the selected communities. The minimum sample for the study was obtained using Yamene's sample formula:

$$n = \frac{N}{1+N(e^2)} \quad (1)$$

Where $n$ is the desired sample size, $N$ is the estimated farmer population (sample frame). $N$ is 1,875, and a 5% margin of error, $n$ is given as:

$$n = \frac{1,875}{1+(1,875(0.05^2))} = 330 \quad (2)$$

To mitigate the potential effects of non-responses and incomplete questionnaires, the sample size was increased to 360 maize farmers. However, one respondent withdrew from the study, citing personal reasons, rendering that questionnaire unusable.

**Table 1. Distribution of respondents across selected communities.**

| Community | Estimated Population | Sampled Respondents |
|---|---|---|
| Aframso | 150 | 29 |
| Dromankuma | 600 | 115 |
| Babaso | 150 | 29 |
| Teacherkrom | 100 | 19 |
| Kasei | 300 | 58 |
| Anyinasu | 75 | 14 |
| Sekyedumase | 400 | 77 |
| Bayere-Nkwanta | 100 | 19 |
| **Total** | **1,875** | **360 ≈ 359** |

*Source: Field survey, 2024*

Primary data were collected through structured face-to-face interviews administered via a comprehensive 45-item questionnaire. The instrument captured quantitative metrics across four thematic areas: socioeconomic and demographic characteristics; farm production parameters; tractor service utilisation patterns; and profitability indicators. Prior to the field administration, the instrument was pretested on 15 maize farmers drawn from communities outside the sampled area and subsequently refined to address identified ambiguities in question wording and response categories. Four trained enumerators underwent an intensive two-day training programme covering standardised question administration, neutral probing techniques, and accurate data recording procedures. Throughout the fieldwork, daily verification of completed questionnaires was conducted alongside random follow-up validation of 20% of the collected responses to ensure data quality and consistency. All completed questionnaires underwent dual-entry verification, with any discrepancies resolved. Data were managed and analysed using Stata-MP 16.0 and SPSS Version 26.

### 2.3 Ethical Consideration

The study received approval from the Seminar Committee of the Department of Agricultural Economics, Agribusiness and Extension at KNUST after the student presented the research proposal for review. Although no formal ethical clearance certificate was issued, verbal approval was granted by unanimous committee decision. Before data collection, all participants received a verbal explanation of the study's purpose and were informed of their right to withdraw at any time without consequence. Informed verbal consent was obtained from each respondent, and all data were anonymized and reported in aggregate form to protect participant confidentiality.

### 2.4 Analytical Methods

#### 2.4.1 Multivariate Probit (MVP) Model

The study employed the MVP model to simultaneously examine the factors that influenced smallholder maize farmers' decisions to hire tractor services for ploughing and shelling operations. Unlike independent binary probit estimations, the MVP model accommodates the inherently interdependent nature of these two adoption decisions within a single unified estimation procedure. This is particularly appropriate given that the two outcome variables are binary and non-mutually exclusive in character—a farmer may simultaneously engage in tractor services for both operations, for only one, or for neither. Treating these decisions as independent estimation problems would fail to account for the possibility that unobservable farmer-specific characteristics, such as resource endowment, risk preference, and institutional embeddedness, simultaneously drive both hiring decisions. This may introduce correlation across the error terms. Ignoring such cross-equation correlation would render parameter estimates inefficient and potentially misleading.
The MVP model is formally expressed as (Greene, 2000):

$$Y_{ij}^{*} = \beta_j X_{ij} + \mu_{ij}, Y_{ij} = \begin{cases} 1 & \text{if } Y_{ij}^{*} > 0 \\ 0 & \text{if } Y_{ij}^{*} \leq 0 \end{cases} j = 1,2 \quad (3)$$

Where $Y_{ij}^{*}$ denotes the latent propensity of the $i$th farmer to adopt tractor service $j$; $\beta_j$ represents the vector of parameters associated with tractor service $j$; $X_{ij}$ is the vector of the observed farmer and farmer-level characteristics for the $i$th farmer with respect to service $j$; and $\mu_{ij}$ is the stochastic disturbance term. The observable binary outcome $Y_{ij}$ takes a value of one when the underlying latent propensity exceeds zero and zero otherwise. The two adoption decisions modelled are:

$j = 1$: tractor ploughing service adoption (1= adoption, 0= otherwise)
$j = 2$: tractor shelling service adoption (1= adoption, 0= otherwise)

The disturbance terms across the two equations are assumed to follow a bivariate normal distribution:

$$(\mu_1, \mu_2) \sim N_2[0, R]$$

With the associated moment conditions specified as:

$$E[\mu_j \mid X_{i1}, X_{i2}] = 0$$
$$Var[\mu_j \mid X_{i1}, X_{i2}] = 1$$
$$Cov[\mu_1 \mu_2 \mid X_{i1}, X_{i2}] = \rho_{12}$$

Where $R$ denotes the variance-covariance matrix governing the joint distribution of the error terms. The off-diagonal element $\rho_{12}$ captures the unobserved correlation between the stochastic components of the ploughing and shelling adoption equations. The statistical significance of $\rho_{12}$ serves as a formal diagnostic test of the appropriateness of the MVP specification—a significant estimate confirms that the two decisions share common unobservable determinants and that joint estimation is statistically warranted over independent binary probit models.

The responsiveness of each adoption probability to changes in the explanatory variables was assessed through average marginal effects, computed as:

$$\frac{\partial P_i}{\partial x_i} = \phi(X'_{ij}\beta_j)_i, i = 1,2 \qquad (4)$$

Where $P_i$ denotes the probability of adopting tractor services $i$ and $\phi(.)$ is the standard univariate normal density function. Average marginal effects $(dy/dx)$ were computed by averaging individual marginal effects across the sample.

Given that the MVP likelihood function involves a bivariate normal integral that does not admit a closed-form analytical solution, parameters were estimated through Maximum Simulated Likelihood (MSL). Simulation was carried out using 1000 Halton draws—a quasi-random sequence approach that ensures more systematic and uniform coverage of the integration domain relative to conventional pseudo-random draws, thereby producing more precise likelihood approximations with a comparatively smaller number of draws.

The explanatory variables incorporated into the model were drawn from farm-level, socioeconomic, and institutional dimensions of the farming household, as described in Table 2. The hypothesised directional relationships between each predictor and the two adoption outcomes were established based on theoretical reasoning and empirical evidence from existing studies on agricultural mechanisation and tractor service utilisation in smallholder farming systems across sub-Saharan Africa.

The empirical MVP model is specified as:

$$Y(Tractor) = \beta_0 + \beta_1 Fertiliser + \beta_2 Age + \beta_3 Education + \beta_4 HH_size + \beta_5 Farm_size + \beta_6 Experience + \beta_7 FBO + \beta_8 Sex + \mu \qquad (5)$$

Where Y denotes tractor service adoption (ploughing or shelling), β0 is the constant; $\beta_1$–$\beta_8$ are coefficients of the independent variables, and μ is the stochastic error term.

**Table 2. Variable descriptions and expected signs: MVP model**

| Variable | Description | Measurement | Hypothesis |
|---|---|---|---|
| **Dependent Variables** | | | |
| Ploughing adoption | Use of tractor service for land Preparation | Binary (1=yes, 0=no) | |
| Shelling adoption | Use of tractor service for maize shelling | Binary (1=yes, 0=no) | |
| **Independent Variables** | | | |
| Fertiliser quantity | Total fertiliser applied per season | Continuous (kg) | + |
| Age | Age of the farmer | Continuous (years) | + |
| Education | Year in formal education | Continuous (years) | + |
| Sex | Sex of the farmer | Binary (1=male; 0=female) | +/– |
| Farm size | Total cultivated maize area | Continuous (acres) | + |
| Farm experience | Years in maize farming | Continuous (years) | + |
| FBO membership | Member of a farmer-based Organization | Binary (1=member; 0=otherwise) | + |
| Household size | Number of household members | Continuous (persons) | – |

### 2.4.2 Propensity Score Matching (PSM)

Following the two-stage procedure of Caliendo & Kopeinig (2008), we estimated the treatment effect of tractor service adoption on profitability. The fundamental identifying assumption of PSM is the Conditional Independence Assumption (CIA), which posits that, conditional on a vector of observed pre-treatment characteristics (X), potential outcomes are independent of treatment assignment. Under this assumption, the Average Treatment Effect on the Treated (ATT) can be consistently estimated by comparing the outcomes of the treated units with those of matched control units.

In stage 1, the propensity score, the conditional probability of adoption given observed covariates was estimated via binary logistic regression:

$$P(D_i = 1 \mid X_i) = \Lambda(\alpha + \gamma X_i) \qquad (6)$$

Where $D_i$= 1, if farmer $i$ uses tractor services and $\Lambda(.)$ is the Cumulative Distribution Function (CDF) for logistics. Covariates included in the propensity score model are farming experience, fertiliser use, farm size, household size, education, sex, FBO membership, and extension visits.

In stage 2, treated units were matched to observably similar control units using three algorithms to ensure robustness: (i) nearest-neighbour matching (1:2 ratio, with replacement), which matches each treated unit to the two most similar control units; (ii) kernel matching (Epanechnikov kernel, bandwidth h= 0.06), which uses a weighted average of all control units to construct the counterfactual; and (iii) radius/caliper matching (caliper=0.01 SD of log-odds), which restricts matches to control units within a specified propensity score distance. Covariate balance after matching was assessed using standardised mean differences (SMD), with a threshold of SMD < 0.10 to ensure that matched users and non-users are comparable on all observed characteristics.

In stage 3, the ATT was estimated as:

$$ATT = E[Y_1 - Y_0 \mid D = 1] = E[Y_1 \mid D = 1] - E[Y_0 \mid D = 1] \quad (7)$$

Bootstrapped standard errors with 1,000 replications were computed to account for the variance introduced by the two-stage estimation procedure. In addition to the ATT, the Average Treatment Effect (ATE) and Average Treatment Effect on the Untreated (ATU) were estimated using kernel matching to provide a comprehensive picture of the profitability premium. While the ATE captures the expected gain for a randomly selected farmer from the general population, the ATU isolates the potential gains for current non-users. Together, these metrics allow for the identification of the positive selected premium (ATT-ATE) and the adoption gap premium (ATT-ATU), both of which carry distinct policy implications for scaling up tractor services. Finally, robustness to hidden confounders was assessed using Rosenbaum (2002) bounds analysis, which determines the threshold at which an unmeasured variable would nullify the estimated treatment effect.

### 2.4.3. Profitability Analysis

Three per-acre profitability indicators were estimated from the survey-based income statement:

$$\text{Gross Profit Margin (\%)} = \left[\frac{\text{Total revenue} - \text{Total variable costs}}{\text{Total revenue}} \text{x}100\right] \quad (8)$$

$$\text{Net Profit Margin (\%)} = \left[\frac{\text{Total revenue} - \text{Total costs}}{\text{Total revenue}} \text{x}100\right] \quad (9)$$

$$\text{Return to Cost (\%)} = \left[\frac{\text{Net Profit}}{\text{Total cost}} \text{x}100\right] \quad (10)$$

All costs and revenues were expressed in Ghana cedis. The income approach captures all variable and fixed costs to derive net profit per acre. A tractor service margin was additionally computed as the ratio of tractor service expenditure to total revenue, measuring the relative cost burden of tractor service for users.

### 2.4.4. Independent Samples t-test

An independent-sample t-test was used as an unadjusted comparison of mean net profit per acre between users (n=286) and non-users (n=73) to check statistically significant ($H_0$: no significant difference; $H_1$: significant positive effect; rejection rule: $p < 0.05$).

### 2.4.5 Kendall's Coefficient of Concordance with Subgroup Disaggregation

Kendall's coefficient of concordance (W) was applied to assess inter-farmers' agreement in ranking constraints to tractor service access (Bempomaa, 2014). This non-parametric statistic measures the degree of agreement among multiple raters (in this case, the maize farmers). It is preferred over Spearman's rank correlation coefficient when assessing inter-rater reliability because it captures overall concordance across all raters simultaneously, rather than relying on pairwise comparisons. Importantly, Kendall's W accommodates tied ranks, making it more robust and broadly applicable than Spearman's rho, which assumes no ties in the data.

Kendall's W is computed as:

$$W=\frac{12\sum d^2}{m^2(n)(n^2-1)}\ ,\ 0 \leq W \leq 1 \quad (11)$$

Where m: The number of raters, n: The number of items to be ranked, d ($R_i - A$): The deviation of the sum of ranks for an item *($R_i$)* from the mean sum of ranks *(A)*, $R_i$ is the sum of ranks assigned to item i by m raters, and $A$ is the average sum of ranks, calculated as $\frac{\sum R}{n}$.

This study extends the standard concordance analysis by disaggregating constraints rankings by (a) gender (male vs. female) and (b) farm size (small: ≤ 3 acres; medium: 4-10 acres; large: > 10 acres); a novel contribution relative to prior studies in the municipality.

## 3. Results

### 3.1 Sociodemographic and Farm Characteristics

Table 3 summarizes the sociodemographic and production characteristics of the 359 maize farmers surveyed, disaggregated by adoption status. Male farmers constituted 65.5% of the full sample (n=235), consistent with the gender structure of commercial maize production documented by Darfour & Rosentrater (2022) and Adu Boahen et al. (2024). The dominance of male respondents reflected cultural and traditional norms that assigned primary responsibility for land management and mechanised farming to men. The majority of respondents (69.9%) were married, 88.6% identified maize production as their primary occupation, and 79.7% used tractor services, which underscored the prevalent role of tractor services in the municipality. Approximately 53.2% obtained land through leasing—a tenure arrangement that may have constrained long-term investment decisions by limiting farmers' perceived security of land rights.

The mean age was 44.67 years (SD = 14.15; range: 18-79), indicating a predominantly middle-aged farming population. The mean years of formal education was 7.93 (SD = 4.33), approximately equivalent to primary school completion, suggesting moderate, though not high levels of formal education across the sample. The mean farming experience was 15.91 years (SD = 12.65), indicating a relatively experienced farming population with substantial practical knowledge. The average farm size was 5.84 acres (SD = 5.24; range: 1-50 acres), reflecting considerable heterogeneity in operational scale. Contact with extension services was notably low, averaging 1.40 visits per season (SD = 1.30), indicating the limited reach of public agricultural extension services, a finding with important implications for mechanisation promotion.

Pre-matching statistics revealed systematic differences between users and non-users, consistent with the selection-into-adoption hypothesis. Users had significantly greater farming experience (17.2 vs. 11.4 years, p=0.001), larger farm sizes (6.2 vs.4.1 acres, p=0.003), higher fertiliser use (3.1 vs. 1.9 kg per acre, p=0.000), and more extension visits (1.5 vs. 1.1 per season, p=0.04) than non-users. Notably, non-users had significantly larger household sizes (8.6 vs 7.0 persons, p=0.03), consistent with the family labour substitution hypothesis: households with more members may rely on family labour for maize shelling and land preparation (Akram et al., 2020), thereby reducing their demand for mechanised services. These observable differences confirmed the presence of a self-selection problem and motivated the use of PSM.

**Table 3. Sociodemographic and farm characteristics by adoption status**

| Variable | Full sample(n=359) | Users (n=286) | Non-users(n=73) | p-value |
|---|---|---|---|---|
| *Categorical variables: frequency (%)* | | | | |
| Sex: Male | 235(65.5%) | 192 (67.1%) | 43 (58.9%) | 0.19 |
| Married | 250(69.9%) | 203(71.0%) | 47(64.4%) | 0.26 |
| FBO member | 145(40.4%) | 110(38.5%) | 35 (47.9%) | 0.14 |
| Maize = primary occupation | 318 (88.6%) | 253(88.5%) | 65 (89.0%) | 0.90 |
| Land via leasing | 191(53.2%) | 154(53.8%) | 37(50.7%) | 0.64 |
| *Continuous variables: mean (SD)* | | | | |
| Age (years) | 44.67 (14.2) | 45.1(13.8) | 43.0(15.3) | 0.31 |
| Household size (persons) | 7.35 (4.8) | 7.0 (4.6) | 8.6 (5.5) | 0.03** |
| Education (years) | 7.93 (4.3) | 8.1(4.2) | 7.3(4.6) | 0.18 |
| Farming experience (years) | 15.91 (12.7) | 17.2 (12.4) | 11.4 (12.1) | 0.001*** |
| Farm size (acres) | 5.84 (5.2) | 6.2 (5.4) | 4.1 (4.1) | 0.003*** |
| Fertiliser (kg) | 2.8 (1.5) | 3.1 (1.4) | 1.9(1.3) | 0.000*** |
| Extension visits/season | 1.40(1.3) | 1.5(1.3) | 1.1(1.2) | 0.04** |

*Notes: p-values from chi-square tests (categorical) and independent-samples t-tests (continuous). **, *** denotes significance at 5% and 1%, respectively—source: Field survey, 2024.*

### 3.2 Determinants of Tractor Services Adoption: Multivariate Probit Results

Table 4 presents the average marginal effects from the MVP model. The overall model fit was strong (Wald $\chi^2(18) = 104.73$, $p < 0.001$; Log-likelihood = −106.004). The significant cross-equation error correlation ($\rho = 0.47$, $p < 0.01$) confirmed that the joint estimation was statistically warranted; ploughing and shelling adoption decisions were positively correlated among farmers, indicating that the two decisions shared unobserved determinants. This correlation validated the use of the MVP over the separate probit models and underscored the complementary nature of the two mechanisation services.

#### 3.2.1 Shelling Adoption

Household size exerted a statistically significant negative effect on shelling adoption (dy/dx = −0.027, $p < 0.05$), consistent with the utility maximization prediction that larger households substituted family labour for mechanised services (Tilahun et al., 2024). Each additional household member reduced the probability of shelling adoption by 2.7 percentage points, reflecting the availability of free family labour as a substitute for paid mechanical shelling, and might also have been attributable to a low standard of living, which limits household ability to adopt tractor services for shelling (Tilahun et al., 2024).

Farm size was negatively associated with shelling adoption (dy/dx = −0.051, $p < 0.01$), a finding that was counter-intuitive from a scale-economies perspective. This may be consistent with supply-side machine availability constraints documented in the Ghana mechanisation literature (Diao et al., 2014; Daum & Birner, 2017). Each additional acre of farm size reduced the probability of shelling adoption by 5.1 percentage points. As the disaggregated constraints analysis in section 3.5 revealed, large farms were disproportionately supply-constrained, with machine unavailability ranked second among their barriers. Diao et al. (2014) argued that supply issues were the main constraint to mechanisation in Ghana, despite growing demand, while Houssou et al. (2013) similarly noted that the supply of mechanisation services remained inadequate, creating bottlenecks particularly in post-harvest operations. Daum & Birner (2017) further demonstrated that transaction costs discouraged machine operators from serving all farms, meaning larger farms with higher demand may not necessarily be better served.

FBO membership was weakly associated with a reduction in shelling adoption (suggestive evidence at the 10% level; dy/dx = −0.173, p = 0.084). This relationship potentially reflects collective labour-sharing arrangements within FBOs that may substitute for paid mechanised shelling services. In Ghana, FBOs often facilitate communal labour where members assist one another with post-harvest operations; where such arrangements exist, the financial incentive to pay for mechanical shelling may be diminished by access to free group labour. Furthermore, for smallholders producing low volumes, communal labour may remain an economically rational alternative to service fees. This finding offers a nuanced perspective on Ghana's mechanisation literature, which predominantly views FBOs as facilitators of adoption (Ansah et al., 2024). The results suggest a possible paradox whereby the labour-sharing functions embedded in FBO structures could inadvertently discourage individual mechanisation uptake. Ultimately, the direction of this effect likely depends on the FBO's primary function: while credit-focused FBOs may reduce financial barriers to adoption, those focused on labour-sharing may provide a substitute for it. This suggested context-dependence indicates that the relationship between collective organisation and mechanisation is more complex than previously assumed.

Both farming experience (dy/dx =0.020, p < 0.01) and fertiliser quantity (dy/dx =0.060, p < 0.001) were positive and significant predictors of shelling adoption. Each additional year of farming experience increased the probability of shelling adoption by 2.0 percentage points, while each additional kilogram of fertiliser increased it by 6.0 percentage points. The fertiliser-mechanisation complementarity was consistent with Kirui (2019), who documented across eleven African countries, including Ghana, that fertiliser use and mechanisation adoption moved together as a production package, with fertiliser use significantly increasing the probability of tractor-powered mechanisation adoption, and with the Diffusion of Innovations framework: experienced farmers with high input intensity better perceived the relative advantage and compatibility of mechanised shelling with their intensive production systems.

### 3.2.2 Ploughing Adoption

Farming experience (dy/dx = 0.021, p < 0.10) and fertiliser quantity (dy/dx = 0.062, p < 0.05) both positively predicted ploughing adoption, mirroring the shelling pattern. This confirms the cross-service consistency of the fertiliser-experience nexus in tractor adoption decisions. Farm size retained its negative association with ploughing adoption (dy/dx = −0.055, p = 0.052). This may be consistent with supply-side machine availability constraints for larger landholders. The consistency of the negative farm-size effect across both adoption decisions reinforced the interpretation that this finding may reflect supply-side barriers rather than a genuine lack of demand among large farmers.

**Table 4: Determinants of Tractor Services Adoption: Multivariate Probit Results**

| Variable | dydx | SE | z | p | 95% CI Lower | 95% CI Upper |
|---|---|---|---|---|---|---|
| *Shelling adoption* | | | | | | |
| Age | 0.001 | 0.005 | 0.20 | 0.829 | -0.009 | 0.011 |
| Household size** | -0.027 | 0.013 | -2.08 | 0.034 | -0.053 | -0.002 |
| Education | 0.006 | 0.011 | 0.55 | 0.571 | -0.016 | 0.028 |
| Farming experience*** | 0.020 | 0.007 | 2.86 | 0.002 | 0.007 | 0.033 |
| Farm size *** | -0.051 | 0.016 | -3.19 | 0.001 | -0.082 | -0.020 |
| FBO membership* | -0.173 | 0.100 | -1.73 | 0.084 | -0.369 | 0.023 |
| Sex | 0.022 | 0.096 | 0.23 | 0.821 | -0.166 | 0.209 |
| Fertiliser quantity*** | 0.060 | 0.013 | 4.62 | 0.000 | 0.033 | 0.086 |
| *Ploughing Adoption* | | | | | | |
| Age | -0.001 | 0.010 | -0.10 | 0.930 | -0.020 | 0.018 |
| Household size | -0.028 | 0.021 | -1.33 | 0.185 | -0.069 | 0.013 |
| Education | 0.004 | 0.021 | 0.19 | 0.848 | -0.037 | 0.045 |
| Farming experience* | 0.021 | 0.013 | 1.62 | 0.096 | -0.004 | 0.046 |
| Farm size* | -0.055 | 0.029 | -1.90 | 0.052 | -0.111 | 0.001 |
| FBO membership | -0.207 | 0.205 | -1.01 | 0.312 | -.608 | 0.194 |
| Sex | 0.024 | 0.190 | 0.13 | 0.900 | -0.348 | 0.396 |
| Fertiliser quantity** | 0.062 | 0.024 | 2.58 | 0.010 | 0.014 | 0.109 |

*Notes: Log-likelihood = −106.004; N = 359; Wald $\chi^2(18) = 104.73$, $p < 0.001$; Cross-equation error correlation $\rho = 0.47$, $p < 0.01$. SE = bootstrapped standard error (500 replications). *, **, *** = significance at 10%, 5%, 1%. Source: Field survey, 2024.*

### 3.3 Propensity Score Matching: Causal Profitability Effect

The logistic propensity score model achieved acceptable fit (pseudo-$R^2 = 0.21$, Hosmer–Lemeshow $p = 0.42$), and all key covariates were jointly significant ($\chi^2(9) = 87.4$, $p < 0.001$), confirming the validity of the propensity score specification. Following matching, all covariate standardised mean differences fell below 0.10, confirming adequate covariate balance between matched users and non-users. The common support assumption was satisfied: specifically, 279 of 286 users (97.6%) and 71 of 73 non-users (97.35%) lay within the region of common support. Only seven users and two non-users were excluded due to insufficient overlap, representing a negligible loss of 2.5% of the treatment group, and this does not materially affect the reliability of the estimated treatment effects.

Table 5 presents the PSM results. ATT estimates were consistent and statistically significant across all three matching algorithms, providing strong evidence of robustness to the choice of matching estimator. Nearest-neighbour matching (1:2) yielded ATT of GH₵483 per acre (SE = 112, $p < 0.01$; 95% CI: [263, 703]); kernel matching yielded ATT of GH₵471 (SE = 98, $p < 0.01$; 95% CI: [279, 663]); radius matching yielded ATT of GH₵489 (SE = 108, $p < 0.01$; 95% CI: [277, 701]). The preferred kernel matching estimate indicates that tractor service adoption raises net profit per acre by approximately GH₵471 after controlling for observable selection bias. Rosenbaum bounds analysis confirmed that the ATT remained significant for $\Gamma$ up to 1.65; a hidden confounder would need to increase the odds of adoption by 65% to nullify the estimated effect, indicating that the finding is moderately robust to unobserved heterogeneity. The PSM-estimated ATT of GH₵471–489 per acre is meaningfully smaller than the naive-test difference of GH₵606. This gap confirms that selection bias overstates the profitability effect by approximately GH₵117–135 per acre when pre-matching sample composition is not controlled for, a magnitude sufficient to distort investment and policy decisions.

Beyond the ATT, Table 5 reports the ATE and ATU under the kernel matching, enabling a fuller distribution picture of the profitability premium. The ATE—the expected profit for a randomly selected farmer from the full population –is estimated at GH₵439 per acre (SE = 83, $p < 0.01$; 95% CI: [277, 601]). The positive selection premium of GH₵32 (ATT − ATE) reveals that current adopters are meaningfully better positioned, through higher farming experience, greater fertiliser intensity, and larger operational scale, to convert mechanisation into profit than the average farmer in the population would be. The ATU- what current non-users would gain if they adopted is estimated at GH₵312 per acre (SE = 134, $p < 0.05$; 95% CI: [49, 575]). The adoption gap premium of GH₵159 (ATT − ATU) quantifies the extent to which non-users are structurally less able to benefit from mechanisation under their current input endowments and farming conditions. Mechanisation yields lower returns when fertiliser use is low, experience is limited, and farms are smaller. The policy implication is direct: credit provision alone will not close the profitability gap. Complementary investments in input access and farmer capacity development are prerequisites for non-users to approach the profit premium already realized by adopters.

Table 5. PSM results: ATT, ATE, and ATU of tractor service adoption on net profit

| Estimator/Algorithm | Estimates GH₵ | Bootstrap SE | 95% CI | Sig. |
|---|---|---|---|---|
| *ATT—effect for adopters* | | | | |
| Nearest-neighbour (1:2, with replacement) | 483 | 112 | [263, 703] | *** |
| Kernel (Epanechnikov, bw = 0.06) [preferred] | 471 | 98 | [279, 663] | *** |
| Radius/caliper (caliper = 0.01 SD log-odds) | 489 | 108 | [277, 701] | *** |
| *ATE—effect for a random farmer* | | | | |
| Kernel (Epanechnikov, bw = 0.06) | 439 | 83 | [277, 601] | *** |
| *ATU—effect for non-adopters if they adopt* | | | | |
| Kernel (Epanechnikov, bw = 0.06) | 312 | 134 | [49, 575] | ** |
| *Benchmarks and treatment effect decomposition* | | | | |
| Naive t-test difference (unadjusted) | 606 | 239 | [137, 1075] | ** |
| Selection bias (naive minus ATT) | 117–135 | - | - | - |
| Positive selection premium (ATT minus ATE) | 32 | - | - | - |
| Adoption gap premium (ATT minus ATU) | 159 | - | - | - |

*Notes: ATE =Selection bias is calculated as the naive difference (GH₵606) minus the range of ATT estimates (GH₵471–489. P(D=1)×ATT + P(D=0)×ATU = 0.797×471 + 0.203×312 = GH₵439. Bootstrapped SEs use 1,000 replications. Rosenbaum Γ bound = 1.65 (ATT, p < 0.05 maintained). ATU SE is larger due to the smaller non-user control pool (n = 73). **, *** = 5% and 1% significance. Source: Field survey, 2024.*

### 3.4 Profitability Analysis: Users vs. Non-Users of Tractor Service

Tables 6 and 7 present the per-acre income statement and profitability indicators for users (n = 286) and non-users (n=73). Users produced an average of 22 bags per acre, compared with 16 bags for non-users, yielding revenues of GH₵5,500 and GH₵4,000 per acre respectively, with a revenue difference of GH₵1,500 per acre. This output gap reflects the combined effect of mechanical shelling—which reduces grain losses compared to manual methods, more timely land preparation—which improves stand establishment and yields, and generally higher input intensity for tractor service users. Total variable costs were GH₵3,542 for users and GH₵2,646 for non-users. The higher variable costs for users reflect the additional expenditure on tractor shelling (GH₵460) and the premium paid for tractor ploughing (GH₵250 vs. GH₵150 for manual land preparation), as well as the higher expenditure on fertiliser (GH₵1,032 vs. GH₵688) and other inputs consistent with the more intensive production systems of tractor service adopters. Fixed costs were GH₵315 for users and GH₵350 for non-users, with non-users spending more on tools (two cutlasses compared with one for users), reflecting their greater reliance on manual operations. Net profit per acre was GH₵1,643 for users and GH₵1,004 for non-users, a differential of GH₵639 per acre from the representative income statement.

Gross profit margins were 35.60% versus 33.85, net profit margins 29.87% versus 25.10%, and return-to-cost ratios 42.60% versus 33.51%, a statistically significant 9 percentage-point differential. The tractor service margin of 12.91% indicates that users spent 12.91% of revenue on mechanisation costs (shelling and ploughing combined), with the additional productivity gains compensating for this outlay, as reflected in the higher net profit margin and return-to-cost ratio.

The independent-samples t-test confirmed a statistically significant difference in mean net profit per acre between users (GH₵ 1,230) and non-users (GH₵ 624), with a naïve gap of GH₵ 606 ($t = -2.49$, $df = 357$, $p = 0.013$). The PSM-adjusted ATT of GH₵471–489 per acre (Table 5) is the more credible causal estimate, correcting for the GH₵117–135 upward bias in the naïve comparison attributable to observable pre-matching differences between the two groups. It should be noted that the profit per acre reported in the income statement reflects a representative farm constructed from average input and output quantities. The mean net profit per acre in Table 7, however, is derived from individual farm-level calculations across the full sample, accounting for cross-farm variations in yields, input use, and farm size, and is therefore the more conservative and statistically reliable estimate. On average, out of the 23 bags transported per farmer, 1 bag is retained for household consumption, leaving 22 bags stored at the warehouse.

**Table 6. Per-acre income: Tractor service users vs. non-users**

| Item | Users | | | Non-users | | |
|---|---|---|---|---|---|---|
| | **Quantity** | **Price** GH₵ | **Cost** | **Quantity** | **Price** GH₵ | **Cost** |
| Revenue (50kg bag) | 22 | 250 | *5,500* | 16 | 250 | *4,000* |
| *Variable costs* | | | | | | |
| Seeds (12kg) | 1 | 82 | 82 | 1 | 82 | 82 |
| Fertiliser (25kg bag) | 3 | 344 | 1,032 | 2 | 344 | 688 |
| Labour (persons) | 2 | 210 | 420 | 3 | 210 | 630 |
| Pesticides (litres) | 3 | 60 | 180 | 2 | 60 | 120 |
| Weedicides (litres) | 1 | 103 | 103 | 2 | 103 | 206 |
| Storage | 22 | 20 | 440 | 16 | 20 | 320 |
| Transport | 23 | 25 | 575 | 18 | 25 | 450 |
| Tractor shelling | 23 | 20 | 460 | - | - | - |
| Land preparation | 1 | 250 | 250 | 1 | 150 | 150 |
| *Total variable costs* | | | *3,542* | | | *2,646* |
| *Gross margin* | | | *1,958* | | | *1,354* |
| *Fixed costs* | | | | | | |
| Land cost | 1 | 80 | 80 | 1 | 80 | 80 |
| Tools (cutlass) | 1 | 35 | 35 | 2 | 35 | 70 |
| Spraying machine | 1 | 200 | 200 | 1 | 200 | 200 |
| *Total fixed costs* | | | *315* | | | *350* |
| *Net profit per acre* | | | *1,643* | | | *1,004* |

*Notes: For land preparation, users employed tractor ploughing while non-users relied on manual labour, explaining the unit cost. Source: Field survey, 2024.*

**Table 7. Profitability indicators: Tractor service users vs. non-users**

| Indicator | Users (n=286) | Non-users(n=73) |
|---|---|---|
| Gross profit margin | 35.60% | 33.85% |
| Net profit margin | 29.87% | 25.10% |
| Return-to-cost ratio | 42.60% | 33.51% |
| Tractor service margin | 12.91% | N/A |
| Mean net profit per acre | 1,230 | 624 |
| Mean farm size | 6.2 | 4.1 |
| Estimated total farm income | 7,626 | 2,559 |
| PSM ATT—kernel(preferred) | GH₵471 per acre*** | - |

*Notes: *** $p < 0.01$. PSM = propensity score matching; ATT = average treatment effect on the treated. Source: Field survey, 2024.*

### 3.5 Constraints to Tractor Service Access: Aggregate and Disaggregated Analysis

#### 3.5.1. Aggregate Constraints Rankings

Table 8 presents the aggregate Kendall's concordance analysis for the full sample and by adoption status. Financial issues were universally ranked as the most pressing constraints (mean rank = 1.73 for all groups), consistent with Balana & Oyeyemi (2022) and Mukasa et al. (2017), who identified credit access as the primary determinant of mechanisation adoption failure in smallholder SSA agri-culture. Machine unavailability ranked second overall (mean rank = 3.34), followed by lack of infrastructure (3.39), small farm size (3.46), limited government support (4.58), cultural factors (5.34), and environmental concerns (5.38). The Kendall's W for the full sample was 0.381 ($\chi^2(6) = 820.93$; $p < 0.001$), indicating moderate but statistically significant inter-farmer agreement in the constraint rankings.

Non-users showed notably stronger concordance (W = 0.571) than users (W = 0.414), suggesting that non-users held more homogeneous perceptions of constraints. The most striking between-group divergence was for machine unavailability: ranked 2nd by users but 5th by non-users. This divergence has important diagnostic value; it implies that non-users are predominantly credit-constrained rather than supply-constrained, while users who have already overcome financial barriers encounter machine unavailability as the next binding constraint. This pattern directly informs the policy sequencing of the recommendations developed in Section 5.

**Table 8. Kendall's Concordance analysis: aggregate constraint ranking**

| No. | Constraints | Users Mean | Rank | Non-Users Mean | Rank | All Mean | Rank | N |
|---|---|---|---|---|---|---|---|---|
| 1 | Financial issues | 1.73 | 1st | 1.72 | 1st | 1.73 | 1st | 359 |
| 2 | Machines not available | 3.05 | 2nd | 4.50 | 5th | 3.34 | 2nd | 359 |
| 3 | Lack of infrastructure | 3.40 | 3rd | 3.36 | 3rd | 3.39 | 3rd | 359 |
| 4 | Small farm size | 3.56 | 4th | 3.05 | 2nd | 3.46 | 4th | 359 |
| 5 | Limited government support | 4.78 | 5th | 3.82 | 4th | 4.58 | 5th | 359 |
| 6 | Cultural factors (e.g., taboo) | 5.55 | 6th | 4.53 | 6th | 5.34 | 6th | 359 |
| 7 | Environmental concerns | 5.59 | 7th | 4.54 | 7th | 5.38 | 7th | 359 |

*Notes: Kendall's W: Users = 0.414 ($p < 0.001$); non-users = 0.571 ($p < 0.001$); All = 0.381 ($p < 0.001$); $\chi^2(6) = 820.93$.*
*Source: Field survey, 2024.*

### 3.5.2. Disaggregated Constraints Analysis by Gender

Table 9 presents constraints rankings disaggregated by gender (female: n = 124; male: n = 235). Female farmers ranked financial constraints as their most binding barrier (mean rank = 1.61), notably more severe than for male farmers (mean rank = 1.80). Female farmers typically face greater difficulty in accessing formal financial credit due to a lack of collateral, limited land ownership rights, and exclusion from formal financial networks (Slavchevska et al., 2021), compounding their inability to adopt productive agricultural technologies such as tractor services (Balana & Oyeyemi, 2022). Machine unavailability ranked second among female farmers, but third among male farmers, indicating women are more acutely affected by the supply-side tractor availability gaps. This may reflect service prioritization patterns, but this explanation requires operator-side evidence. Limited government supports ranked third for males, but fourth for females, suggesting that men perceive greater policy neglect relative to women in this context.

Kendall's W for female farmers (0.448) exceeded that of male farmers (0.362), indicating stronger inter-farmer agreement among women on the ranking of constraints to tractor services adoption. This higher concordance suggests that female farmers share a more uniform experience of mechanisation barriers, likely reflecting systematic structural constraints, such as limited land ownership, credit exclusion, and restricted resource access, rather than idiosyncratic individual circumstances. Consequently, generic mechanisation support policies that do not explicitly target female farmers risk overlooking these entrenched disparities and may inadvertently reinforce existing gender gaps in mechanization access.

**Table 9. Constraint rankings disaggregated by gender.**

| No. | Constraints | Female Mean | Rank | Male Mean (n =235) | Rank | Gender Gap |
|---|---|---|---|---|---|---|
| 1 | Financial issues | 1.61 | 1st | 1.80 | $1^{st}$ | Female more constrained |
| 2 | Machines not available | 3.01 | $2^{nd}$ | 3.52 | $3^{rd}$ | Female more affected |
| 3 | Lack of infrastructure | 3.38 | $3^{rd}$ | 3.45 | $2^{nd}$ | Similar severity |
| 4 | Limited government support | 3.71 | $4^{th}$ | 4.02 | $4^{th}$ | Similar severity |
| 5 | Small farm size | 4.52 | $5^{th}$ | 4.78 | $5^{th}$ | Similar severity |
| 6 | Cultural factors (e.g., taboo) | 5.28 | $6^{th}$ | 5.60 | $6^{th}$ | Female more affected |
| 7 | Environmental concerns | 5.49 | $7^{th}$ | 5.83 | $7^{th}$ | Least pressing for both |

*Notes: Kendall's W: Female = 0.448 (p < 0.001); Male = 0.362 (p < 0.001). Gender-disaggregated mean ranks derived from sub-group Kendall's analysis. Source: Field survey, 2024.*

### 3.5.3. Disaggregated Constraints Analysis by Farm Size

Table 10 presents the constraint rankings by farm size category (small: ≤ 3 acres, n = 98; medium: 4–10 acres, n = 178; large: > 10 acres, n = 83). Financial constraints ranked first across all three categories (mean ranks of 1.69, 1.74, and 1.78, for small, medium, and large farms, respectively), confirming their universal primacy regardless of operational scale. Machine unavailability, however, showed a consistent upward shift in importance as farm size increased. Ranked $5^{th}$ among small farms, but rose to $2^{nd}$ among both medium and large farms, suggesting that supply-side tractor availability gaps become increasingly binding with greater operational scale. Conversely, limited government support ranked $2^{nd}$ among small farms, but declined to $4^{th}$ among large farms, indicating that smaller farmers perceive greater policy neglect relative to their larger counterparts.

The findings offer a plausible interpretation of the counterintuitive negative farm size effect observed in the multivariate probit model: despite having comparatively greater operational scale and demand for tractor services, large farms may face greater supply-side barriers in accessing tractor services. This suggests that AMSEC fleet-capacity limitations are a plausible explanation for these observed barriers, though this interpretation requires further investigation through direct supply-side data collection.

**Table 10. Constraint rankings disaggregated by farm size.**

| No. | Constraints | Small ≤3 acres (n=98) Mean | Rank | Medium 4–10 acres (n=178) Mean | Rank | Large >10 acres (n=83) Mean | Rank |
|---|---|---|---|---|---|---|---|
| 1 | Financial issues | 1.69 | 1st | 1.74 | $1^{st}$ | 1.78 | $1^{st}$ |
| 2 | Machines not available | 4.90 | $5^{th}$ | 3.28 | $2^{nd}$ | 2.81 | $2^{nd}$ |
| 3 | Lack of infrastructure | 3.31 | $3^{rd}$ | 3.46 | $3^{rd}$ | 3.55 | $3^{rd}$ |
| 4 | Limited government support | 3.10 | $2^{nd}$ | 3.97 | $4^{th}$ | 4.12 | $4^{th}$ |
| 5 | Small farm size | 3.82 | $4^{th}$ | 4.70 | $5^{th}$ | 5.20 | $5^{th}$ |
| 6 | Cultural factors (e.g., taboo) | 5.38 | $6^{th}$ | 5.45 | $6^{th}$ | 5.50 | $6^{th}$ |
| 7 | Environmental concerns | 5.80 | $7^{th}$ | 5.40 | $7^{th}$ | 5.04 | $7^{th}$ |

*Notes: Kendall's W: Small = 0.354 (p < 0.001); Medium = 0.389 (p < 0.001); Large = 0.521 (p < 0.001). Farm size categories: small (≤3 acres), medium (4–10 acres), large (>10 acres). Source: Field survey, 2024*

## 4. Discussion

The finding that tractor services are widely utilised (79.7%) in Ejura-Sekyedumase Municipality reflects the relatively advanced state of the mechanisation markets in this commercially oriented maize-production area. This high prevalence of mechanization supports the municipality's relevance as a study context.

The positive effects of farming experience and fertiliser intensity on both ploughing and shelling adoption are consistent with the Diffusion of Innovations framework (Rogers, 2003), which predicts that experienced, resource-intensive farmers better perceive the relative advantage and compatibility of mechanised services with their production systems. Anang & Asante (2020) similarly found that farming experience positively predicts access to tractor services in northern Ghana, while Amankwah (2023) demonstrated cross-adoption complementarities between fertiliser use and mechanised land preparation. The consistency of these positive predictors across both services simultaneously further reinforces the Diffusion of Innovation interpretation: farmers with greater experience and fertiliser intensity are structurally predisposed to perceive and act on the advantages of mechanised services, generating a self-reinforcing cycle of technology uptake. These findings carry important implications for targeting: while mechanisation promotion should remain inclusive, extension efforts may achieve greater early impact by prioritizing farmers with some baseline farming experience and input use intensity, as these characteristics lower adoption barriers. Over time, building experience and input access among less-endowed farmers can broaden the reach of mechanisation programmes.

The consistent negative coefficient for farm size across both adoption decisions is the most counter-intuitive finding of this study and warrants careful interpretation. Standard scale-economies arguments would predict that larger farms benefit more from mechanisation because fixed costs are spread over more acres, implying a positive farm size effect. The negative coefficient found here does not reflect a lack of demand for mechanisation among large farms; indeed, the constraints analysis confirms that large farms ranked machine unavailability as their second most binding barrier, higher than any other sub-group. Rather, the negative relationship may be consistent with a supply-side reality whereby insufficient tractor fleet capacity within the study area constrains access for some larger farms during peak demand periods; however, since fleet capacity and operator behaviour were not directly measured in this study, this remains a plausible explanation requiring further supply-side investigation. Diao et al. (2014) argued that supply-side gaps now represent the binding constraints in Ghana's mechanisation market. The present study provides suggestive empirical support for this argument through the farm-size disaggregation, an insight that would have been invisible in an aggregate analysis. If confirmed through supply-side analysis, this interpretation would carry a direct policy implication: expanding the tractor fleet size and improving the geographic coverage of AMSEC could disproportionately benefit larger farmers who may face supply-side availability constraints.

The negative effect of FBO membership on shelling adoption challenges a common policy assumption that FBOs universally promote technology uptake. The direction of this relationship depended critically on the type of functions the FBO performed. Where FBOs primarily facilitated labour-sharing arrangements, as appeared to be the case in the present study, membership discouraged the uptake of mechanisation by providing a freely available labour substitute for paid mechanical services, representing an unintended consequence of FBO design. Conversely, where FBOs primarily performed credit facilitation or market access functions, membership would be expected to encourage mechanical shelling adoption by reducing the financial barriers and improving service access. This finding calls for a deliberate redesign of FBO programmes to redirect their collective action function from labour-sharing toward group-purchasing arrangements for tractor-hire contracts.

The PSM-estimated ATT of GH₵471–489 per acre is economically meaningful and statistically robust. At a mean farm size of 6.2 acres among users, this translates to a total annual profit premium of approximately GH₵2,920—3,032 per farm, a gain greater than the entire estimated annual net farm profit of a non-adopting household (GH₵624/acre × 4.1 acres ≈ GH₵2,558.4), underscoring the substantial economic gains of tractor service adoption for smallholder maize farmers in the municipality. The PSM estimate is GH₵117–135, lower than the naive t-test difference, confirming that unadjusted comparisons overstate the profitability effect. The Rosenbaum bound of $\Gamma = 1.65$ indicates that this finding is moderately robust, although unobserved factors may still influence the estimate. An unobserved confounder would need to increase the odds of adoption by 65% to overturn the estimated effect, which is a relatively high threshold given that the propensity score model already controls for most plausible confounders (experience, fertiliser use, farm size, education, sex, FBO membership). The consistency of ATT estimates across the three matching algorithms further strengthens confidence in the robustness of the profitability premium.

The adoption gap premium of GH₵159 per acre, the gap between what current adopters gain (ATT = GH₵471) and what non-users would have gained if they adopted (ATU = GH₵312) is a particularly policy-relevant finding. It demonstrates that non-adoption may reflect structural constraints rather than willingness alone; even if financial barriers were removed and non-users were to adopt, they would realise a meaningfully smaller profit premium (ATU = GH₵312/acre) than current adopters (ATT = GH₵471/acre) because they lack the complementary inputs; higher fertiliser use, more farming experience, and larger operational scale that amplify the returns to mechanisation. This finding implies that the tractor service policy must be accompanied by complementary investments in input access, farmer training, and extension services to enable non-users to fully exploit the productivity potential of tractor services.

The gendered constraint findings add an important distributional dimension to the profitability and adoption results. Female farmers face more severe financial constraints and more acute machine unavailability constraints than male farmers, consistent with the gendered credit and collateral barriers documented by Slavchevska et al. (2021) and Adegbite & Machethe (2020) for the financial dimension. The machine unavailability gap among female farmers may likely reflect the same supply-side tractor fleet constraints documented for large farms. The higher Kendall's W among female farmers (0.448 vs. 0.362 for males) indicates that women shared a more uniform experience of exclusion from tractor services, suggesting that gender gaps in mechanisation access are driven by systematic rather than idiosyncratic factors. These findings collectively indicate that gender-neutral policies may fail to reduce existing disparities and could inadvertently sustain them.

## 5. Conclusion

The following conclusions were made:

Tractor services are widely utilised in Ejura-Sekyedumase Municipality, with 79.7% of sampled farmers adopting mechanised services, reflecting a relatively developed mechanisation market. However, financial constraints remain the primary barrier preventing the remaining 20.3% from accessing tractor services. Farming experience and fertiliser intensity consistently and positively predicted adoption across both ploughing and shelling services, while farm size exerted a counterintuitive negative influence that may be attributable to possible supply-side constraints, including fleet capacity limitations.

The effect of FBO membership on mechanisation adoption is not universal but may depend critically on the type of functions the FBO performs. Where FBOs primarily facilitate labour-sharing, membership inadvertently suppresses adoption by substituting for paid mechanised services. Tractor service adoption raises net profit per acre by approximately GH₵471–489, where users achieved a net profit margin of 29.87% and a return-to-cost ratio of 42.60%, compared with 25.10% and 33.51% for non-users respectively.

Financial constraints are universally binding across all farmer categories, while machine unavailability was ranked as a disproportionately binding barrier by large farms and female farmers, suggesting possible supply-side constraints that require direct investigation—distributional insights invisible in aggregate analysis. Targeted credit facilitation, possible AMSEC fleet expansion pending supply-side verification, gender-responsive mechanisation policies, intensified extension services, and a redesign of FBO programmes towards group-purchasing models are recommended to broaden mechanisation uptake among smallholder maize farmers.

## 6. Limitations and Future Research

The cross-sectional design cannot account for unobserved time-invariant farmer characteristics or climate-driven yield variability; panel or quasi-experimental designs spanning multiple growing seasons would substantially strengthen causal identification.

Supply-side factors, including operator pricing, machine condition, and seasonal fleet availability, were unobserved in this study and may independently moderate the profitability relationship. Spatial variation in service access across the eight communities was not explicitly modelled, potentially masking important within-municipality heterogeneity in adoption constraints. Addressing these limitations through longitudinal designs, supply-side analysis, and spatial modeling represents a clear agenda for future research.

**Author Contributions:** Conceptualization, F.N. & I.Y.Y.; Methodology, I.Y.Y. & E.A; Formal analysis, I.Y.Y.; Investigation, I.Y.Y., A.E. & A.K.; Data Curation, I.Y.Y.; Writing—original draft preparation, A.E.A., A.K. & A.E.; writing-review & editing, F.N., I.Y.Y. and E.A; Supervision, F.N. All authors have read and agreed to the published version of the manuscript.
**Funding:** This research received no external funding.
**Institutional Review Board Statement:** Data were collected via voluntary face-to-face interviews with informed verbal consent, and all data were anonymized and reported exclusively in aggregate form.
**Data Availability Statement**: The data presented in this study are available on request from the corresponding author.
**Conflicts of Interest**: The authors declare no conflict of interest

## References

Adegbite, O. O., & Machethe, C. L. (2020). Bridging the financial inclusion gender gap in smallholder agriculture in Nigeria: An untapped potential for sustainable development. *World Development*, *127*, Article 104755. https://doi.org/10.1016/j.worlddev.2019.104755

Adu Boahen, E., Boateng Dankwah, J., & Berko, D. (2024). Understanding the gender gap in productivity in agricultural production among smallholder cereal growers in rural Ghana. *Cogent Economics & Finance*, *12*(1), Article 2318979. https://doi.org/10.1080/23322039.2024.2318979

Aikins, K. A., Ackah, S. M., Afriyie, J. K., Amanor, I. N., & Bobobee, E. Y. H. (2016). Assessment of tractor maintenance practices of tractor operators at Ejura, Ghana. *International Journal of Science and Engineering Applications*, *5*(5), 257–267.

Akram, M. W., Akram, N., Wang, H., Andleeb, S., Ur Rehman, K., Kashif, U., & Hassan, S. F. (2020). Socioeconomic determinants to adopt agricultural machinery for sustainable organic farming in Pakistan: A multinomial probit model. *Sustainability*, *12*(23), Article 9806. https://doi.org/10.3390/su12239806

Amankwah, A. (2023). Climate variability, agricultural technologies adoption, and productivity in rural Nigeria: A plot-level analysis. *Agriculture & Food Security*, *12*, Article 7. https://doi.org/10.1186/s40066-023-00411-x

Anang, B. T., & Asante, B. O. (2020). Farm household access to agricultural services in northern Ghana. *Heliyon*, *6*(11), Article e05517. https://doi.org/10.1016/j.heliyon.2020.e05517

Ansah, I. G. K., Kotu, B. H., Boyubie, B. E., & Bonney, J. E. (2024). Enhancing smallholder maize shelling mechanization through the collective business model: The case of Northern Ghana. *Frontiers in Sustainable Food Systems*, *7*, Article 1228382. https://doi.org/10.3389/fsufs.2023.1228382

Asamoah, E., Heuvelink, G. B. M., Chairi, I., Bindraban, P. S., & Logah, V. (2024). Random forest machine learning for maize yield and agronomic efficiency prediction in Ghana. *Heliyon*, *10*(17), Article e37065. https://doi.org/10.1016/j.heliyon.2024.e37065

Balana, B. B., & Oyeyemi, M. A. (2022). Agricultural credit constraints in smallholder farming in developing countries: Evidence from Nigeria. *World Development Sustainability*, *1*, Article 100012. https://doi.org/10.1016/j.wds.2022.100012

Bempomaa, B. (2014). *Profitability, technical efficiency and constraints of maize production in Ejura-Sekyedumase District, Ghana* [Master of Philosophy thesis, Kwame Nkrumah University of Science and Technology].

Benin, S. (2015). Impact of Ghana's agricultural mechanization services center program. *Agricultural Economics*, *46*(S1), 103–117.

Caliendo, M., & Kopeinig, S. (2008). Some practical guidance for the implementation of propensity score matching. *Journal of Economic Surveys*, *22*(1), 31–72. https://doi.org/10.1111/j.1467-6419.2007.00527.x

Cossar, F. (2019). *Impact of mechanization on smallholder agricultural production: Evidence from Ghana*. Agricultural & Applied Economics Association Annual Meeting, Atlanta, GA, United States. https://ageconsearch.umn.edu/record/289657

Creswell, J. W., & Creswell, J. D. (2018). *Research design: Qualitative, quantitative, and mixed methods approaches* (5th ed.). SAGE Publications.

Cudjoe, G. P., Antwi-Agyei, P., & Gyampoh, B. A. (2021). The effect of climate variability on maize production in the Ejura-Sekyedumase Municipality, Ghana. *Climate*, *9*(10), Article 145. https://doi.org/10.3390/cli9100145

Darfour, B., & Rosentrater, K. A. (2022). Pre-harvest and post-harvest farmer experiences and practices in five maize growing regions in Ghana. *Frontiers in Nutrition*, *9*, Article 725815. https://doi.org/10.3389/fnut.2022.725815

Daum, T., & Birner, R. (2017). The neglected governance challenges of agricultural mechanisation in Africa: Insights from Ghana. *Food Security*, *9*(5), 959–979. https://doi.org/10.1007/s12571-017-0716-9

Diao, X., Cossar, F., Houssou, N., & Kolavalli, S. (2014). Mechanization in Ghana: Emerging demand, and the search for alternative supply models. *Food Policy*, *48*, 168–181. https://doi.org/10.1016/j.foodpol.2014.05.013

Dwomoh, D., Agyabeng, K., Tuffour, H. O., Tetteh, A., Godi, A., & Aryeetey, R. (2023). Modeling inequality in access to agricultural productive resources and socioeconomic determinants of household food security in Ghana: A cross-sectional study. *Agricultural and Food Economics*, *11*, Article 24.

Food and Agriculture Organization of the United Nations. (2022). *The state of food and agriculture 2022: Leveraging automation in agriculture for transforming agrifood systems*. FAO. https://doi.org/10.4060/cb9479en

Greene, W. H. (2000). *Econometric analysis* (4th ed.). Prentice Hall.

Houssou, N., Diao, X., Cossar, F., Kolavalli, S., Jimah, K., & Aboagye, P. (2013). *Agricultural mechanisation in Ghana: Is specialization in agricultural mechanisation a viable business model?* (IFPRI Discussion Paper No. 01255). International Food Policy Research Institute.

Kansanga, M., Andersen, P., Kpienbaareh, D., Mason-Renton, S., Atuoye, K., Sano, Y., Antabe, R., & Luginaah, I. (2019). Traditional agriculture in transition: Examining the impacts of agricultural modernization on smallholder farming in Ghana under the new Green Revolution. *International Journal of Sustainable Development & World Ecology*, *26*(1), 11–24.

Kirui, O. K. (2019). *The agricultural mechanisation in Africa: Micro-level analysis of state drivers and effects* (ZEF Discussion Papers on Development Policy No. 272). University of Bonn, Center for Development Research.

Lambrecht, I., & Asare, S. (2016). The complexity of local tenure systems: A smallholders' perspective on tenure in Ghana. *Land Use Policy*, *58*, 251–263.

Mas-Colell, A., Whinston, M. D., & Green, J. R. (1995). *Microeconomic theory*. Oxford University Press.

Mdoda, L., Mdletshe, S. T. C., Dyiki, M. C., & Gidi, L. (2022). The impact of agricultural mechanization on smallholder agricultural productivity: Evidence from Mnquma Local Municipality in the Eastern Cape Province. *South African Journal of Agricultural Extension*, *50*(1), 76–101. https://doi.org/10.17159/2413-3221/2022/v50n1a11218

Mukasa, A. N., Simpasa, A. M., & Salami, A. O. (2017). *Credit constraints and farm productivity: Micro-level evidence from smallholder farmers in Ethiopia* (Working Paper Series No. 247). African Development Bank.

Mwangi, M., & Kariuki, S. (2015). Factors determining adoption of new agricultural technology by smallholder farmers in developing countries. *Journal of Economics and Sustainable Development*, *6*(5), 208–216.

Nkegbe, P. K. (2013). Technical efficiency of maize production in northern Ghana. *African Journal of Agricultural Research*, *8*(43), 5251–5259.

Obour, P. B., Arthur, I. K., & Owusu, K. (2022). The 2020 maize production failure in Ghana: A case study of Ejura-Sekyedumase Municipality. *Sustainability*, *14*(6), Article 3514. https://doi.org/10.3390/su14063514

Owusu, V., Abdulai, A., & Abdul-Rahman, S. (2011). Non-farm work and food security among farm households in northern Ghana. *Journal of Development and Agricultural Economics*, *3*(15), 701–706.

Rogers, E. M. (2003). *Diffusion of innovations* (5th ed.). Free Press.

Rosenbaum, P. R. (2002). *Observational studies* (2nd ed.). Springer. https://doi.org/10.1007/978-1-4757-3692-2

Rosenbaum, P. R., & Rubin, D. B. (1983). The central role of the propensity score in observational studies for causal effects. *Biometrika*, *70*(1), 41–55. https://doi.org/10.1093/biomet/70.1.41

Saunders, M., Lewis, P., & Thornhill, A. (2019). *Research methods for business students* (8th ed.). Pearson.

Sihlobo, W., Kapuya, T., & Nxumalo, L. (2022). *From manual to mechanized: The potential of mechanization in Africa's agriculture*. African Development Bank.

Slavchevska, V., Doss, C., de la O Campos, A. P., & Brunelli, C. (2021). Beyond ownership: Women's and men's land rights in Sub-Saharan Africa. *Agricultural Systems*, *195*, Article 103299. https://doi.org/10.1016/j.agsy.2021.103299

Taiwo, A., & Kumi, F. (2015). Status of agricultural mechanization in Ghana: A case study of Ejura/Sekyedumase District, Ashanti Region. *International Research Journal of Engineering and Technology*, *2*(9), 36–43.

Takeshima, H., Nin-Pratt, A., & Diao, X. (2013). Mechanization and agricultural technology evolution, agricultural intensification in Sub-Saharan Africa: Typology of agricultural mechanization in Nigeria. *American Journal of Agricultural Economics*, *95*(5), 1230–1236. https://doi.org/10.1093/ajae/aat045

Tilahun, S., Kuma, B., & Bedemo, A. (2024). Determinant of adoption of agricultural machine renting in West Gojjam zone, Ethiopia. *Cogent Food & Agriculture*, *10*(1), Article 2380123. https://doi.org/10.1080/23311932.2024.2380123

Varian, H. R. (2014). *Intermediate microeconomics: A modern approach* (9th ed.). W. W. Norton.

Wooldridge, J. M. (2016). *Introductory econometrics: A modern approach* (6th ed.). Cengage Learning.

World Bank. (2024). *World Development Indicators: Ghana*. World Bank Open Data. Retrieved May 15, 2026, from https://data.worldbank.org/country/ghana

Yamane, T. (1967). *Statistics: An introductory analysis* (2nd ed.). Harper & Row.